\documentclass[aps,epsf,showpacs,floatfix,twocolumn]{revtex4}
\usepackage{amsmath}
\usepackage{amssymb}
\usepackage{subfigure}
\usepackage{graphicx}
\usepackage[usenames]{color}

\begin{document}

\title{Two-component gap solitons with linear interconversion}
\author{Sadhan K. Adhikari$^{1}$ and Boris A. Malomed$^{2}$}
\affiliation{$^1$Instituto de F\'{\i}sica Te\'{o}rica, UNESP -- S\~{a}o Paulo State
University, 01.405-900 S\~{a}o Paulo, S\~{a}o Paulo, Brazil \\
$^2$Department of Physical Electronics, School of Electrical Engineering,
Tel Aviv University, Tel Aviv 69978, Israel}

\begin{abstract}
We consider one-dimensional solitons in a binary Bose-Einstein
condensate with linear coupling between the components, trapped in
an optical-lattice potential. The inter-species and intra-species
interactions may be both repulsive or attractive. Main effects
considered here are spontaneous breaking of the symmetry between
components in symmetric and antisymmetric solitons, and spatial
splitting between the components. These effects are studied by
means of a variational approximation and numerical simulations.
\end{abstract}

\maketitle

\textit{Introduction:} The existence of quasi-one-dimensional (1D) solitons
in Bose-Einstein condensates (BECs) has been demonstrated in well-known
works \cite{solitons}, which used the Feshbach-resonance (FR)\ technique to
switch the repulsion between atoms into attraction. It was also predicted
\cite{GSprediction}, and then demonstrated in the experiment \cite{Markus},
that an optical-lattice (OL) potential may support gap solitons (GSs) in
finite bandgaps of the OL-induced spectrum.

Experiments are possible too in a binary BEC, created as a mixture of two
hyperfine states of the same atom \cite{binary}. In that case, the
inter-species interaction may also be controlled by FR \cite{inter-Feshbach}%
. It was proposed to use this setting for the creation of \textit{symbiotic}
solitons \cite{symbiotic}, in which the FR-induced inter-species attraction
overcomes the intra-species repulsion. Two-component \textit{symbiotic GSs}
were predicted too \cite{Arik}. In particular, the attraction between two
self-repulsive species may lead, counter-intuitively, to \emph{spatial
splitting} between GSs formed in each species \cite{Warsaw}, which is
explained by a\ negative effective mass of the GS. A general two-component
model including an OL potential acting on both species was considered in
Ref. \cite{we}, with inter-species repulsion and intra-species repulsion or
attraction. Several types of GSs were reported there: symmetric and
asymmetric ones, unsplit or split complexes (in which, respectively, centers
of the two components coincide or are separated), and solitons classified as
\textit{intra-gap} or \textit{inter-gap} ones, with chemical potentials of
the two components belonging to the same or different bandgaps.

Our objective here is to extend the consideration to a different physical
situation, when the interaction between the two components trapped in the OL
includes linear coupling, i.e., interconversion between the species. The
linear interconversion between hyperfine atomic states can be induced by a
resonant electromagnetic field \cite{interconversion}, leading to the
prediction of various coupled-mode dynamical effects, such as Josephson
oscillations \cite{Josephson}.

On the contrary to the previously studied symbiotic solitons \cite%
{symbiotic,Arik,Warsaw,we}, chemical potentials of the two components of
stationary patterns in the linearly-coupled binary BEC must be equal. This
condition, obviously, favors symmetric states, i.e., ones with identical
components. On the other hand, nonlinear repulsion will push them aside.
Thus, the competition between the linear coupling and nonlinear
inter-component repulsion gives rise to a shift of the
miscibility-immiscibility transition in a binary BEC, or in a superfluid
Fermi gas trapped in a parabolic potential \cite{students}; however,
manifestations of such competition in terms of solitons were not considered
before, and this is one of objectives of the present work. On the other
hand, the interplay of the intra-component self-repulsion or self-attraction
and linear interconversion gives rise to \textit{spontaneous symmetry
breaking} (SSB) in 1D \cite{Arik1} and 2D \cite{Arik2} binary solitons. The
SSB manifests itself in the destabilization of symmetric or antisymmetric
solitons (in the case of the self-attraction or repulsion, respectively),
and the emergence of stable asymmetric states, with different numbers of
atoms in the components. The SSB in two-component GSs was also predicted in
optics, in terms of dual-core fiber Bragg gratings \cite{BG}.

In this work, we use the variational approximation (VA) and numerical
simulations to demonstrate that, in the presence of the linear
interconversion of components, both the SSB and spatial-splitting transition
in binary-BEC solitons trapped in the OL potential take a simple form, which
can be represented by universal diagrams in the parameter space.

\textit{The model:} We consider a binary BEC loaded into
``cigar-shaped" trap, combined with the OL potential acting in the
longitudinal direction. The system of coupled Gross-Pitaevskii
equations for
the two wave functions, $\phi _{1}$ and $\phi _{2}$, can be written as \cite%
{symbiotic,Arik,Warsaw,we}:%
\begin{eqnarray}
i\left( \phi _{j}\right) _{t} &=&-(1/2)\left( \phi _{j}\right) _{xx}+{\ }%
\left( g|\phi _{j}|^{2}+g_{12}\left\vert \phi _{3-j}\right\vert ^{2}\right)
\phi _{j}  \notag \\
&&-V_{0}\cos \left( 2x\right) \phi _{j}-\kappa \phi _{3-j},\quad j=1,2.
\label{q1}
\end{eqnarray}%
The scaled coordinate, time, OL strength, linear-interconversion rate, and
nonlinearity coefficients are related to their counterparts measured in
physical units: $x\equiv \left( \pi /L\right) x_{\mathrm{ph}}$, $t\equiv
\left( \pi /L\right) ^{2}\left( \hbar /m\right) t_{\mathrm{ph}}$, $%
V_{0}\equiv \left( L/\pi \hbar \right) ^{2}m(V_{0})_{{\mathrm{ph}}}$, and $%
\kappa =\left( L/\pi \right) ^{2}\left( m/\hbar \right) \kappa _{\mathrm{ph}%
} $, and $\left\{ g,g_{12}\right\} \equiv \left( Lm\omega _{\perp }\mathcal{N%
}/\pi \hbar \right) \left\{ a,a_{12}\right\} $, where $L$, $m$, $\omega
_{\perp }$, and $\mathcal{N}$ are the OL period, atomic mass,
transverse-trapping frequency, and total number of atoms in both species. We
define $V_{0}$ and $\kappa $ to be positive, while $a,a_{12}>0$ and $<0$
correspond to the nonlinear repulsion and attraction, respectively. With $t$
replaced by propagation distance $z$ and $g_{12}=0$, Eqs. (\ref{q1}) may
also be interpreted as a model for the spatial evolution of optical signals
in a planar dual-core waveguide, equipped with a transverse grating of
strength $V_{0}$ \cite{Arik1}.

Stationary solutions to Eqs. (\ref{q1}) are looked for in the usual form, $%
\phi _{1,2}(x,t)=\exp \left( -i\mu t\right) u_{1,2}(x)$, with a common
chemical potential, $\mu $. Real functions $u_{1,2}(x)$ obey the following
equations and normalization:
\begin{equation}
\mu u_{j}+u_{j}^{\prime \prime }/2-{\ }\left( gu_{j}^{2}+{\ g_{12}}%
u_{3-j}^{2}\right) u_{j}+V_{0}\cos \left( 2x\right) u_{j}+\kappa u_{j}=0,
\label{stationaryx}
\end{equation}%
$N_{1}+N_{2}=2,~N_{j}\equiv \int_{-\infty }^{+\infty }u_{j}^{2}(x)dx$, which
can be derived from the Lagrangian,%
\begin{eqnarray}
&&L=\int_{-\infty }^{+\infty }\biggr\{\sum_{j=1,2}\biggr[\mu
u_{j}^{2}-(u_{j}^{\prime })^{2}/2+V_{0}\cos (2x)u_{j}^{2}  \notag \\
&&-gu_{j}^{4}/2\biggr]-g_{12}u_{1}^{2}u_{2}^{2}+2\kappa u_{1}u_{2}\biggr\}%
dx-2\mu .  \label{L}
\end{eqnarray}%
Solitons may exist if $\mu $ falls into the semi-infinite gap (SIG)\ or
finite bandgaps of the linear spectrum of system (\ref{stationaryx}), which
was found in Ref. \cite{Arik1}.

\textit{The variational approximation}. To predict solitons with a compact
unsplit profile and, generally, different numbers of atoms in the
components, $N_{1}\neq N_{2}$, we adopt the Gaussian ansatz \cite{VA},
\begin{equation}
u_{j}^{\left( \mathrm{unspl}\right) }(x)=\sigma ^{j}\pi ^{-1/4}\sqrt{N_{j}/w}%
e^{-x^{2}/(2w^{2})},\quad j=1,2,  \label{ansatz}
\end{equation}%
where, for $N_{1}=N_{2}$, $\sigma =+1$ and $-1$ correspond to symmetric and
antisymmetric solitons, respectively: $u_{1}^{\left( \mathrm{unspl}\right)
}=\pm u_{2}^{\left( \mathrm{unspl}\right) }$. Free parameters in ansatz (\ref%
{ansatz}) are norms $N_{1,2}$ of the components and their common width $w$
(in the presence of the linear coupling, one may assume equal widths of the
components, even if their amplitudes are different \cite{VA,Arik1}). The
substitution of ansatz (\ref{ansatz}) in the Lagrangian yields%
\begin{eqnarray}
L &=&\mu \left( N-2\right) -N/\left( 4w^{2}\right) +V_{0}e^{-w^{2}}N  \notag
\\
&+&2\sigma \kappa \nu -\left[ g\left( N^{2}-\nu ^{2}\right) +g_{12}\nu ^{2}%
\right] /\left( 2\sqrt{2\pi }w\right) ,  \label{LagrAB}
\end{eqnarray}%
where $N\equiv N_{1}+N_{2}$, $\nu \equiv \sqrt{2N_{1}N_{2}}$. In this
notation, the asymmetry parameter of the soliton is%
\begin{equation}
\epsilon \equiv \left( N_{1}-N_{2}\right) /\left( N_{1}+N_{2}\right) =\sqrt{%
1-\nu ^{2}/2}.  \label{epsB}
\end{equation}

The first variational equation, $\partial L/\partial \mu =0$, recovers the
normalization adopted above, $N=2$. The other equations, $\partial
L/\partial w=\partial L/\partial \nu =0$ and $\partial L/\partial N=0$, yield%
\begin{gather}
\sqrt{2\pi }\left( 1-4V_{0}w^{4}e^{-w^{2}}\right) +\left[ g\left( 1+\epsilon
^{2}\right) +g_{12}\left( 1-\epsilon ^{2}\right) \right] w=0,  \label{wB} \\
2\sqrt{\pi }\kappa w=\sigma \left( g_{12}-g\right) \sqrt{1-\epsilon ^{2}},
\label{BB}
\end{gather}%
and $\mu =\left( 4w^{2}\right) ^{-1}-V_{0}e^{-w^{2}}+\sqrt{2/\pi }gw^{-1}$.
Note that Eq. (\ref{BB}) has no solutions for $\sigma \left( g_{12}-g\right)
<0$, which means that stationary asymmetric states with identical signs of
the two wave functions ($\sigma =+1$) do not exist if the repulsion between
the species is weaker than the intrinsic repulsion in each of them, or,
alternatively, if the inter-species attraction is stronger than its
intra-species counterpart. Just the opposite is true for states with
different numbers of atoms and opposite signs of the wave functions, $\sigma
=-1$. The equation for width $w_{0}$ of symmetric and antisymmetric states
is obtained from Eq. (\ref{wB}) by setting $\epsilon =0$ in it,
\begin{equation}
4V_{0}w_{0}^{3}\exp \left( -w_{0}^{2}\right) -w_{0}^{-1}=\left(
g+g_{12}\right) /\sqrt{2\pi },  \label{symm}
\end{equation}%
while Eq. (\ref{BB}) should be ignored in this case, as its derivation
implied a deviation from the symmetry.

An issue of major interest is to predict a critical value of the
linear-coupling constant, $\kappa =\kappa _{\mathrm{bif}}$, at which the SSB
bifurcation occurs, i.e., solutions to Eqs. (\ref{wB}) and (\ref{BB}) with
infinitesimal $\epsilon $ split off from the solutions with $\epsilon =0$.
In this case, Eq. (\ref{wB}) again reduces to (\ref{symm}), but Eq.\ (\ref%
{BB}) should not be omitted, yielding%
\begin{equation}
\kappa _{\mathrm{bif}}=\sigma \left( g_{12}-g\right) /\left( 2\sqrt{\pi }%
w_{0}\right) .  \label{bif}
\end{equation}%
%
%
%

The VA may also be applied predict the \textit{splitting border }in the case
of the repulsive nonlinearity. To approximate the onset of the splitting, we
use the ansatz introduced in Ref. \cite{we} (in the absence of the linear
coupling),%
\begin{equation}
u_{i}^{\mathrm{(spl)}}(x)=\frac{\sigma ^{n}\sqrt{N}}{\pi ^{1/4}\sqrt{w}}%
\left[ 1\pm bx-\frac{1}{2}w^{2}\left( b^{2}-x^{2}\right) \right] e^{-{x^{2}}%
/({2w^{2}})},  \label{A}
\end{equation}%
where the separation between the centers of the components is $\Delta
x\approx 2bw^{2}$ for small $b$. The substitution of this ansatz in
Lagrangian (\ref{L}) yields
\begin{eqnarray}
&&L=-\frac{1}{2w^{2}}+2V_{0}e^{-w^{2}}-\frac{g+g_{12}}{\sqrt{2\pi }w}%
+2\sigma \kappa \\
&&-2b^{2}\left( 2V_{0}w^{4}e^{-w^{2}}-\frac{g_{12}w}{\sqrt{2\pi }}+\sigma
\kappa w^{2}\right) ,
\end{eqnarray}%
and the additional variational equation, $\partial L/\partial \left(
b^{2}\right) =0$, predicts the splitting condition:%
\begin{equation}
g_{12}=2\sqrt{2\pi }V_{0}w_{0}^{3}e^{-w_{0}^{2}}+\sigma \sqrt{2\pi }\kappa
w_{0}.  \label{g}
\end{equation}

\textit{Numerical results.} Numerical solutions of Eq. (\ref{q1}) were
obtained by means of the real-time propagation using the split-step
Crank-Nicolson algorithm \cite{we}, with spatial and temporal steps $0.025$
and $0.0002$, respectively. The simulations were run until the solution
would settle down into a stationary localized state. This method of
obtaining stationary solutions guarantees their stability.

\begin{figure}[tbp]
\begin{center}
\includegraphics[width=.7\linewidth]{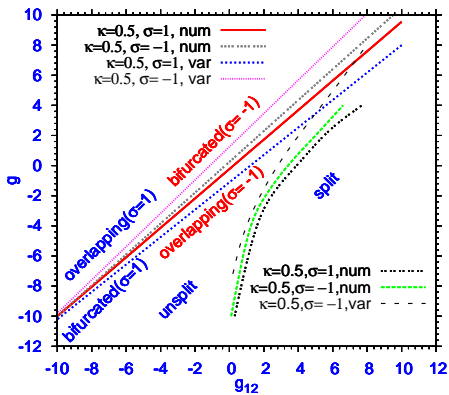}
\end{center}
\caption{(Color online) Four curves on the left represent lines of
the spontaneous symmetry breaking of symmetric ($\protect\sigma
=+1$) and antisymmetric ($\protect\sigma =-1$) solitons, for
$V_{0}=5$. Three curves on the right designate the onset of the
spatial splitting between the components of symmetric and
antisymmetric solitons. The results are obtained from the
variational approximation (VA) and numerical simulations (num).
Labels ``overlapping/bifurcated" and ``unsplit/split" indicate,
respectively, areas of the existence of stable
symmetric/asymmetric solitons, and of those with
coinciding/separated centers of the two components.} \label{fig1}
\end{figure}

The most fundamental information about the transition between different
types of two-component solitons is provided by the SSB and splitting borders
in the plane of interaction coefficients $\left( g_{12},g\right) $, for
fixed coupling constant $\kappa $. A generic example of such a diagram is
presented in Fig. \ref{fig1} for $V_{0}=5$ and $\kappa =0.5$ (in physical
units, the latter corresponds to interconversion time $\sim 50$ $\mu $s, for
atoms of $^{7}$Li trapped in the OL with period $L\sim 1~\mu $m; the range
of values of $\left\vert g,g_{12}\right\vert $ displayed in Fig. \ref{fig1}
corresponds to the solitons built of up to $\sim 10^{5}$ atoms). The
variational results displayed in the diagram are obtained from Eqs. (\ref%
{bif}) and (\ref{g}), using a numerical solution of Eq. (\ref{symm}) for $%
w_{0}$. In the range of $0.1\lesssim \kappa \lesssim 1$, and for other
relevant values of $V_{0}$, the diagram keeps essentially the same form as
in Fig. \ref{fig1}.

The diagram covers both positive (repulsive) and negative (attractive)
values of $g_{12}$ and $g$, the SSB line in quadrants $g_{12},g>0$ and $%
g_{12},g<0$ going, respectively, through families of GSs and regular
solitons (the latter ones belong to the SIG). It is worthy to stress that
the broken-symmetry areas are located on \emph{opposite sides} of the SSB
lines for $\sigma =+1$ and $-1$, i.e., for the the symmetric and
antisymmetric solitons. The VA version of the splitting line for $\sigma =+1$
is not included in Fig. \ref{fig1}, as in this case SSB happens prior to the
onset of the splitting, while Eq. (\ref{g}) was derived from the VA assuming
the unbroken symmetry. Note also that the SSB and splitting lines do not
intersect, and there is no overlap between the stability areas of different
types of the two-component solitons, i.e., the model does not give rise to a
bistability.

The splitting lines displayed in Fig. \ref{fig1} are similar to those
reported in Ref. \cite{we} for the model with $\kappa =0$ (nevertheless, the
difference is that $\kappa \neq 0$ makes \textit{complete separation} of the
two components of the soliton impossible, unlike the situation with $\kappa
=0$). On the other hand, all the results concerning the SSB lines have no
previously published counterparts, as the symmetry breaking in regular
and/or gap-mode solitons was not studied before in models featuring the
competition between the linear coupling and nonlinear interaction between
the species.

\begin{figure}[tbp]
\begin{center}
$%
\begin{array}{cc}
\includegraphics[width=.48\linewidth]{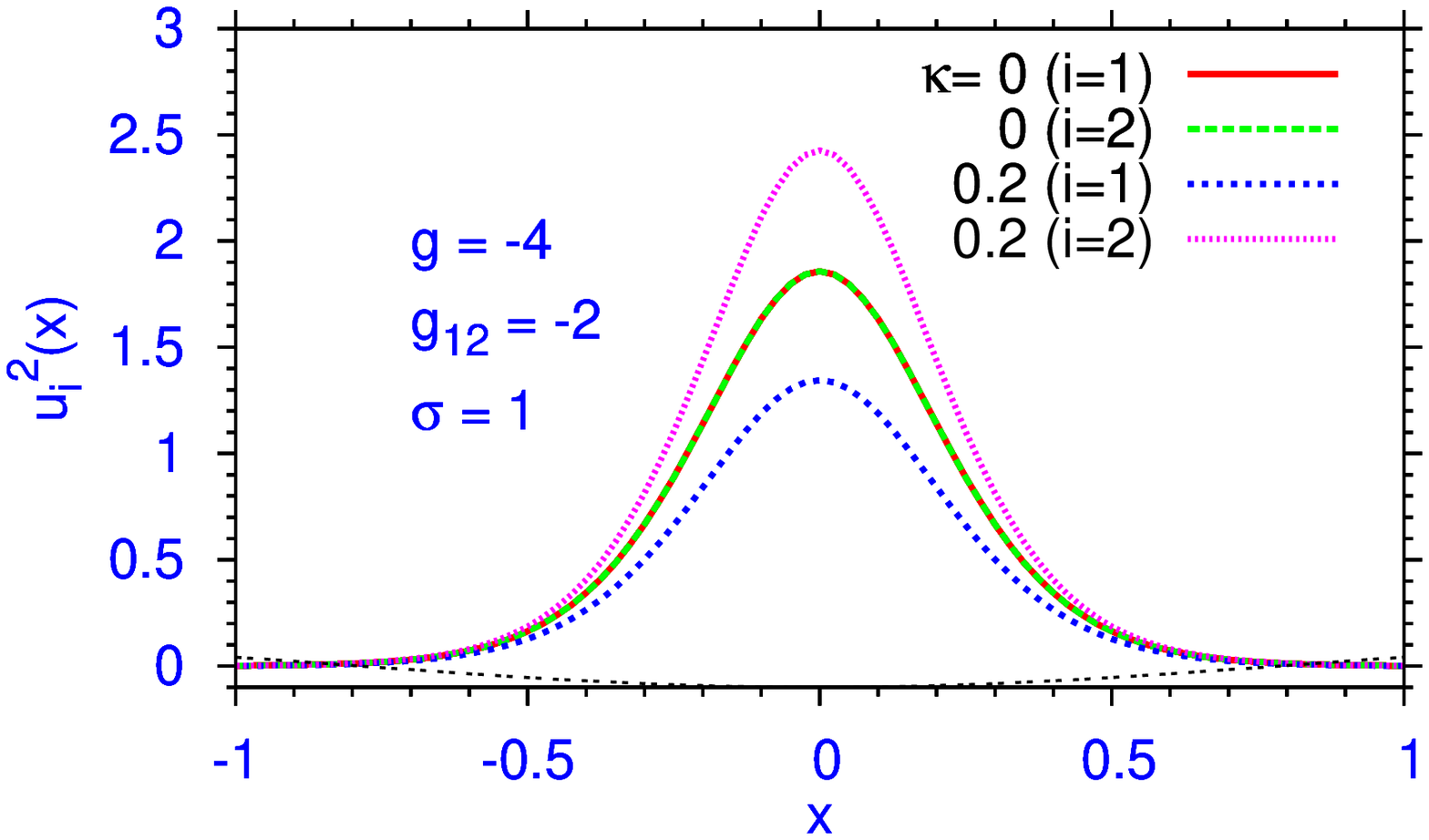} & %
\includegraphics[width=.48\linewidth]{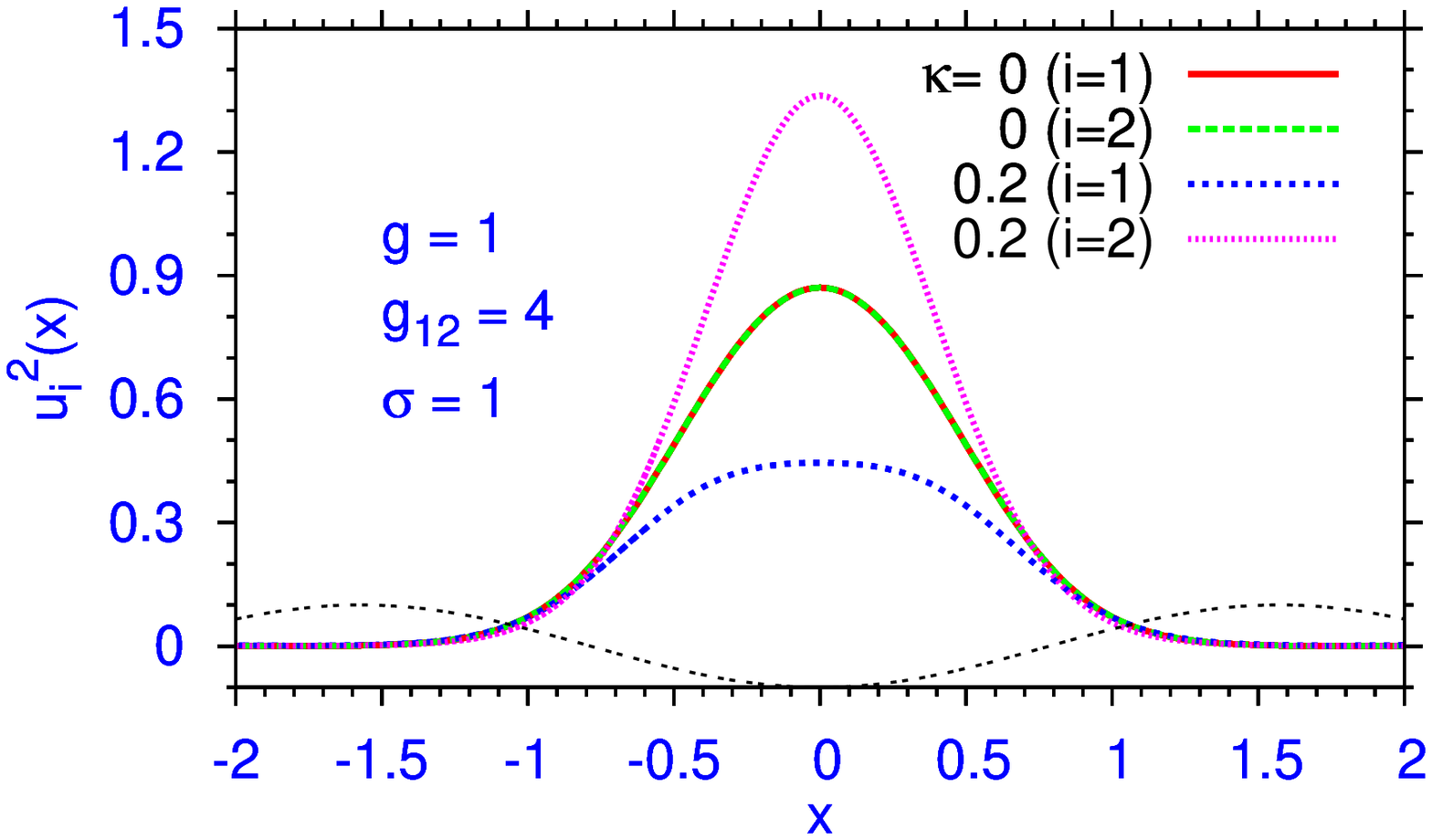} \\
\mathrm{(a)} & \mathrm{(b)} \\
\includegraphics[width=.48\linewidth]{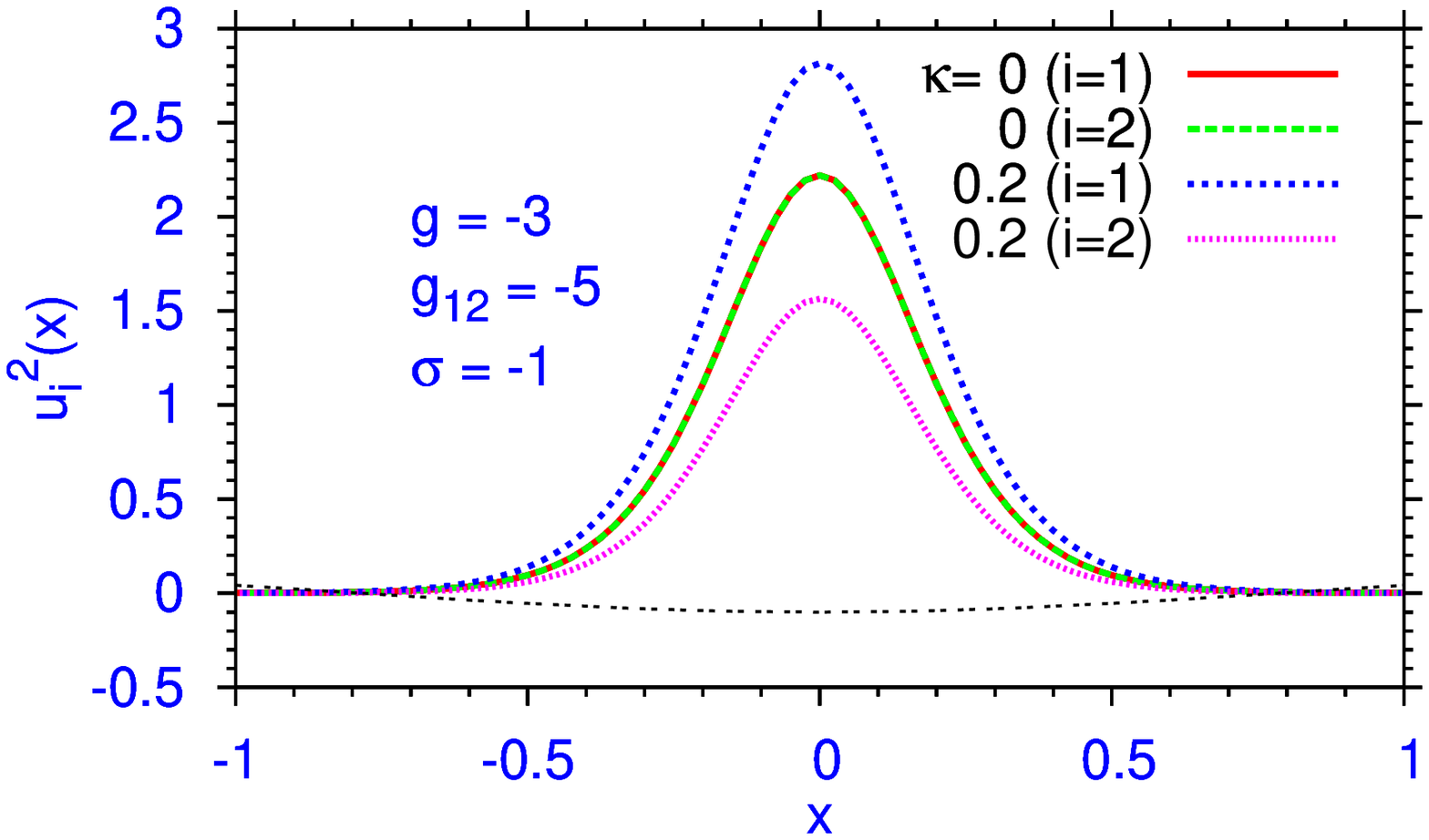} & %
\includegraphics[width=.48\linewidth]{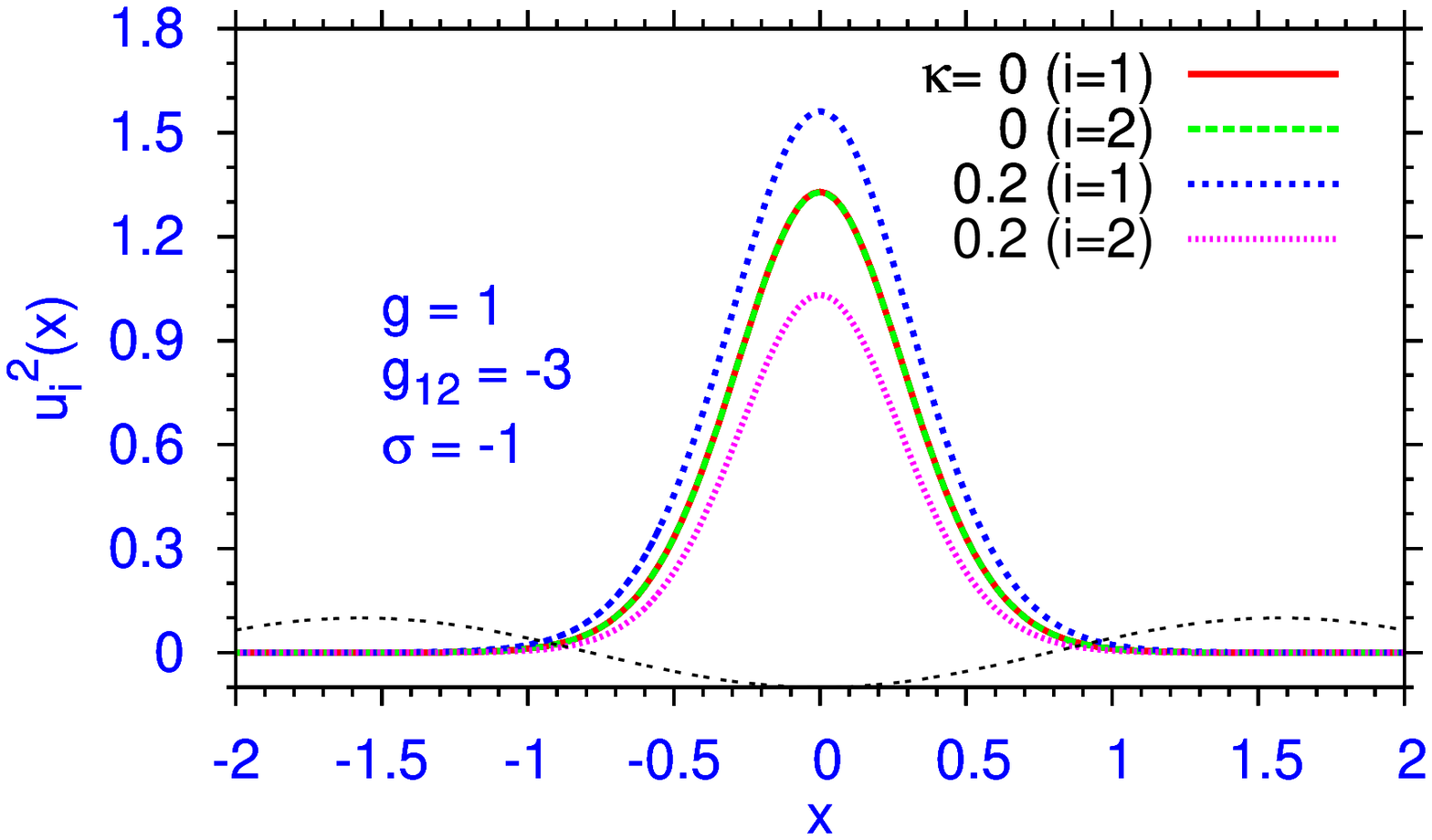} \\
\mathrm{(c)} & \mathrm{(d)} \\
\includegraphics[width=.48\linewidth]{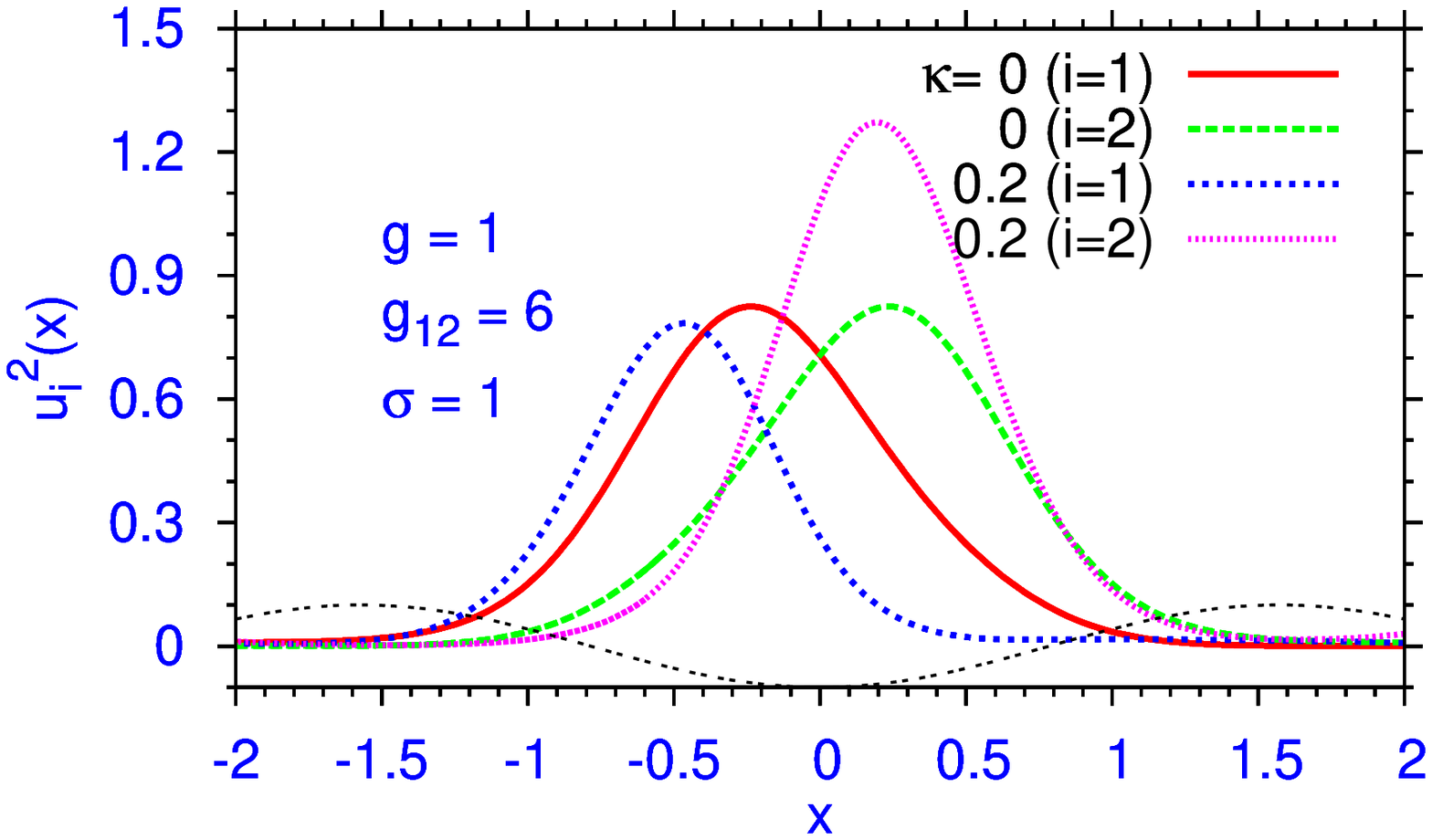} & %
\includegraphics[width=.48\linewidth]{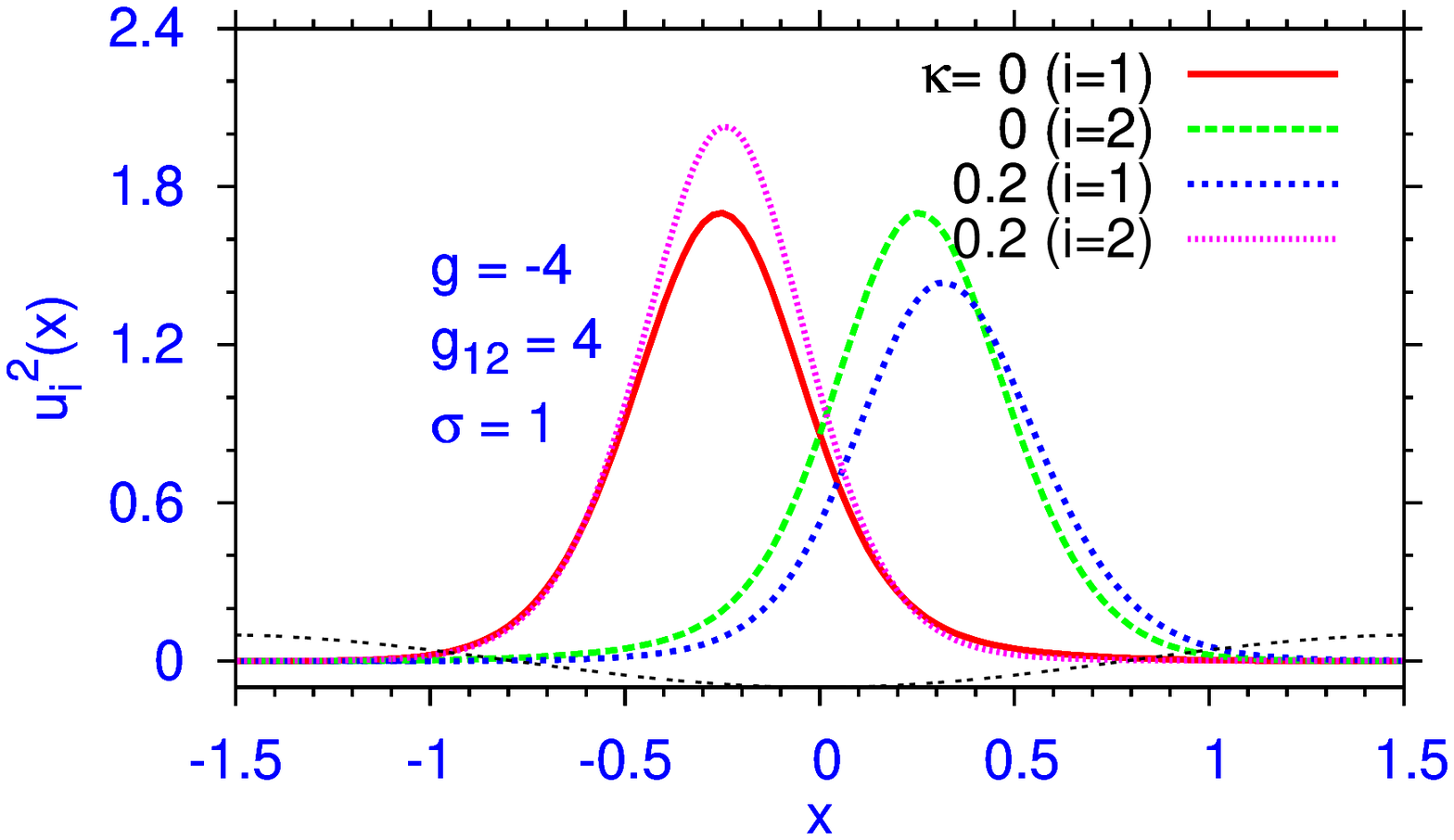} \\
\mathrm{(e)} & \mathrm{(f)}%
\end{array}%
$%
\end{center}
\caption{(Color online) Examples of stable symmetric/asymmetric and
unsplit/split solitons, as obtained from the numerical solution for
different combinations of the attractive and repulsive interactions. The
thin dashed curve depicts the OL potential, with $V_{0}=5$.}
\label{fig2}
\end{figure}

Figure \ref{fig2} displays typical examples of SSB in symmetric and
antisymmetric solitons [panels (a), (b) and (c), (d), respectively], and of
the spatial splitting combined with SSB [(e), (f)]. To stress the role of
the linear coupling in inducing the transition to the asymmetric shapes, the
panels also include the case of $\kappa =0$. For all unsplit solitons, the
shapes predicted by the VA, both symmetric and asymmetric ones, are
virtually identical to the numerically found shapes shown in panels (a)-(d).

\begin{figure}[tbp]
\begin{center}
{\includegraphics[width=\linewidth]{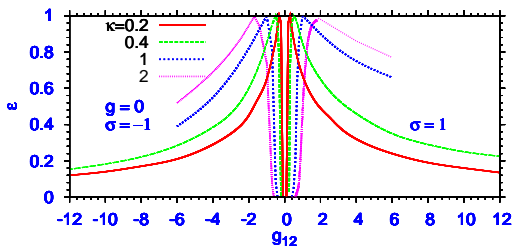}} 
\end{center}
\caption{(Color online) The asymmetry of the solitons, $\protect\epsilon $,
defined as per Eq. (\protect\ref{epsB}), versus the nonlinear-coupling
coefficient, $g_{12}$, as found from the numerical solution for $V_{0}=5,$ $%
g=0$, and $\protect\sigma =1$ (four lines on the right side)  or 
$\protect\sigma =-1$ (four lines on the left side).}
\label{fig3}
\end{figure}

The case when the intra-species nonlinearity is switched off ($g=0$), and
the solitons exist only due to the inter-species interactions, is of
particular interest. The evolution of families of stable asymmetric GSs with
the increase of $\left\vert g_{12}\right\vert $, as obtained from the
numerical solution in this case (both for the attraction, $g_{12}<0$, and
repulsion, $g_{12}>0$, when the solitons can be found, severally, only with $%
\sigma =-1$ or $\sigma =+1$), is shown in Fig. \ref{fig3}, for different
values of linear coupling $\kappa $. In particular, the initial abrupt
increase of $\epsilon $ with the growth of $g_{12}>0$ is the manifestation
of the SSB in the unsplit GS. After achieving a maximum of $\epsilon $ very
close to $1$, the asymmetric GS undergoes the splitting transition, which
eventually leads to the gradual decrease of the effective asymmetry between
the spatially separating components. In the case of $g_{12}<0$, the solitons
(in this case, they belong to the SIG), also attain a maximum of $\epsilon $
very close to $1$. Further increase of $\left\vert g_{12}\right\vert $ leads
to a transition to solitons of the symbiotic type \cite{symbiotic} with
reduced asymmetry. These results, as well as those summarized in Fig. \ref%
{fig1}, do not have previously published counterparts either.

\begin{figure}[tbp]
\begin{center}
{\includegraphics[width=0.7\linewidth]{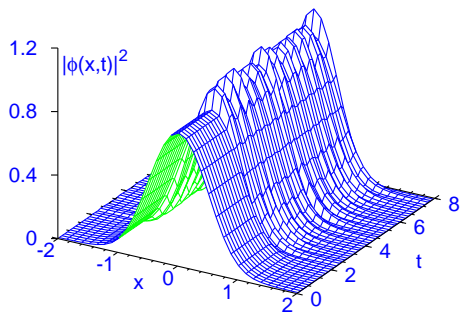}}
\end{center}
\caption{(Color online) The time evolution of the symmetric gap soliton
found at $\protect\sigma =1,$ $g=4,$ $g_{12}=2,$ $\protect\kappa =0.2,$ $%
V_{0}=5$, after as $g$ suddenly dropped from $4$ to $1$. Here,
$\phi_1=\phi_2\equiv\phi(x,t)$.} \label{fig4}
\end{figure}

As mentioned above, the numerical procedure adopted in this work generates
only stable soliton solutions. However, unlike the ordinary SIG-based
solitons, GSs do not represent the BEC ground state \cite{GSprediction,Markus}%
, i.e., they are, strictly speaking, metastable objects. For this reason, it
is relevant to test their stability against strong perturbations.
Simulations demonstrate that the GSs are, in fact, very robust objects. For
instance, sudden drop of the self-repulsion coefficient from $g=4$ to $g=1$,
which can be easily implemented by means of FR, does not destroy the GS, see
Fig. \ref{fig4}. Similarly, sudden application of a kick to the soliton (not
shown here) gives rise to its oscillations around a local minimum of the OL
potential, but does not destroy it either.

\textit{Conclusion}. In this work, we have extended the model of binary BEC
trapped in the OL potential \cite{we} by including a the linear
interconversion between the two species. Using the VA and numerical
simulation, we have identified two internal transitions in two-component
solitons, both regular and gap-mode ones: spontaneous symmetry breaking, and
spatial splitting between the components.

This work was supported, in a part, by FAPESP and CNPq (Brazil).


\begin{thebibliography}{99}
\bibitem{solitons} K. E. Strecker \textit{et al}.,
Nature \textbf{417}, 150 (2002); L. Khaykovich \textit{et al.},
Science \textbf{256}, 1290 (2002);
S. L. Cornish, S. T. Thompson and C. E. Wieman, Phys. Rev. Lett. \textbf{96}%
, 170401 (2006).

\bibitem{GSprediction} O. Zobay \textit{et al.},
Phys. Rev. A \textbf{59}, 643 (1999); 
A. Trombettoni and A. Smerzi, Phys. Rev. Lett. \textbf{86}, 2353 (2001);
B. B. Baizakov, V. V. Konotop, and M. Salerno, J. Phys. B \textbf{35}, 51015
(2002); P. J. Y. Louis \textit{et al.},
Phys. Rev. A \textbf{67}, 013602 (2003).

\bibitem{Markus} B. Eiermann \textit{et al.},
Phys. Rev. Lett. \textbf{92}, 230401 (2004).


\bibitem{binary} C. J.\ Myatt \textit{et al.},%
Phys. Rev. Lett. \textbf{78}, 586 (1997); D. M.\ Stamper-Kurn \textit{et al.}%
, 
Phys.\ Rev.\ Lett.\ \textbf{80}, 2027 (1998).

\bibitem{inter-Feshbach} A.\ Simoni \textit{et al.}
, Phys.\ Rev.\ Lett.\ \textbf{90}, 163202 (2003).

\bibitem{symbiotic} V. M. P\'{e}rez-Garc\'{\i}a and J. B. Beitia, Phys. Rev.
A \textbf{72}, 033620 (2005); S. K. Adhikari, Phys. Lett. A 346, 179 (2005);
Phys. Rev. A \textbf{72}, 053608 (2005); \textbf{70}, 043617 (2004); \textbf{%
76}, 053609 (2007); J. Phys. A \textbf{40}, 2673 (2007).

\bibitem{Arik} A. Gubeskys, B. A. Malomed, and I. M. Merhasin, Phys. Rev. A
\textbf{73}, 023607 (2006).


\bibitem{Warsaw} M. Matuszewski, B. A. Malomed, and M. Trippenbach, Phys.
Rev. A. \textbf{76}, 043826 (2007).

\bibitem{we} S. K. Adhikari and B. A. Malomed,
Phys. Rev. A 77, 023607 (2008).

\bibitem{interconversion} R. J. Ballagh, K. Burnett, and T. F. Scott, Phys.
Rev. Lett. \textbf{78}, 1607 (1997).

\bibitem{Josephson} J. Williams \textit{et al}.,%
Phys. Rev. A \textbf{59}, R31 (1999); P. \"{O}hberg and S. Stenholm, Phys.
Rev. A \textbf{59}, 3890 (1999); D. T. Son and M. A. Stephanov, Phys. Rev. A
\textbf{65}, 063621 (2002); S. D. Jenkins and T. A. B. Kennedy, Phys. Rev. A
\textbf{68}, 053607 (2003).

\bibitem{students} M. I. Merhasin, B. A. Malomed, and R. Driben,
J. Phys. B \textbf{38}, 877
(2005); S. K. Adhikari and B. A. Malomed, Phys. Rev. A 74, 053620 (2006).

\bibitem{Arik1} A. Gubeskys and B. A. Malomed,
Phys. Rev. A \textbf{75}, 063602 (2007).

\bibitem{Arik2} A. Gubeskys and B. A. Malomed,
Phys. Rev. A \textbf{76}, 043623 (2007); M. Matuszewski, B. A. Malomed, and
M. Trippenbach,
Phys. Rev. A \textbf{75}, 063621 (2007); M. Trippenbach \textit{et al}.%
, 
Phys. Rev. A 78, 013603 (2008).

\bibitem{BG} W. C. K. Mak, B. A. Malomed, and P. L. Chu,
J. Opt. Soc. Am. B \textbf{15}, 1685
(1998);
Phys. Rev. E \textbf{69}, 066610 (2004); Y. J. Tsofe and B. A. Malomed,
\textit{ibid}. \textbf{75}, 056603 (2007); S. Ha and A. A. Sukhorukov, J.
Opt. Soc. Am. B \textbf{25}, C15 (2008).

\bibitem{VA} V. M. P\'{e}rez-Garc\'{\i}a \textit{et al.},
Phys. Rev. A \textbf{56} 1424 (1997); B. A. Malomed, Progress in Optics
\textbf{43}, 71 (2002).


\bibitem{Merhasin} A. Gubeskys, B. A. Malomed, and I. M. Merhasin,
Stud. Appl. Math. \textbf{115}, 255
(2005).
\end{thebibliography}
\end{document}